\documentclass[11pt,a4paper]{article}
\usepackage{jcappub}

\title{The virial theorem in Eddington-Born-Infeld gravity}

\author[a]{Noelia S. Santos} \author[a]{ and Janilo Santos}
	
\affiliation[a]{Departament of Theoretical and Experimental Physics, Federal University of Rio Grande do Norte, 59072-970 Natal - RN, Brazil}

\emailAdd{noelia@dfte.ufrn.br}

\emailAdd{janilo@dfte.ufrn.br}

\abstract{We consider the possibility that the Eddington-Born-Infeld (EBI) modified gravity provides an alternative explanation for the mass discrepancy in clusters of galaxies. For this purpose we derive the modified Einstein field equations, finding an additional "geometrical mass" term which provides an effective contribution to the gravitational binding energy. Using some approximations and assumptions for weak gravitational fields, and taking into account the collisionless relativistic Boltzmann equation, we derive a generalized version of the virial theorem in the framework of EBI gravity. We show that the "geometrical mass" term may account for the well known virial mass discrepancy in  clusters of galaxies. We also derive the velocity dispersion relation for galaxies in the clusters, which could provide an efficient method for testing EBI gravity from astrophysical observations.}


\arxivnumber{1506.04569}

\begin{document}
	\maketitle


\section{Introduction}


One of the major challenges of modern astrophysics is certainly to explain the mass gravitationally bounded in structures like galaxies and clusters of galaxies.  The incompatible behavior of the rotation curves of spiral galaxies with the theoretical prediction of Newton's gravity~\cite{vera} - to which the Einstein's theory reduces at this scale - as well as the mass discrepancy in clusters of galaxies~\cite{navarro462,delpopolo,Newman}, are usually explained by postulating the existence of a cold pressureless fluid, called dark matter, which interacts only gravitationally  (see~\cite{Martins} for explanatory reviews on the dark matter properties). Despite of many attempts~\cite{mann}, the nature of dark matter is still unknown and the only convincing evidence for its existence is gravitational.
In clusters of galaxies the strong bending of light indicates that there is a lot of matter heavily concentrated in a central region of the cluster. As is well known, the total mass of a cluster can be estimated at least in two ways. First, by considering the motions of the member galaxies of the cluster and using the virial theorem to estimate the virial mass $M_{\mbox{\scriptsize{v}}}$. Second, by adding the mass of each individual galaxy member of the cluster, the total baryonic mass $M_{\mbox{\scriptsize{b}}}$ is determined. It is found that $M_{\mbox{\scriptsize{v}}}$ is much greater than $M_{\mbox{\scriptsize{b}}}$, with typical values of $M_{\mbox{\scriptsize{v}}}/M_{\mbox{\scriptsize{b}}}\sim 20 - 30$ \cite{sal,anna,binney}. This discrepancy is usually attributed to the existence of dark matter. Very important for gravitationally bounded systems, like the ones found in Astrophysics, the virial theorem is, however, dependent on the gravitational theory. Hence, an alternative way to attack the problem of discrepancy of the mass of clusters of galaxies is modifying the theory of gravity (for a thoroughly comprehensive review on modified gravity and its consequences see~\cite{Clifton}). This issue has recently received a lot of attention and several versions of the virial theorem are found in the scientific literature by using modified gravity such as metric $f(R)$ theory~\cite{lobo}, Palatini $f(R)$ theory~\cite{sefi}, metric-Palatini hybrid gravity~\cite{capozziello}, brane-world models~\cite{harko} and DGP-inspired $L(R)$ gravity~\cite{DGP}.
In this paper we study the virial theorem in matter-gravity coupling modifications based on the so-called Eddington-Born-Infeld (EBI) gravity theory~\cite{banados, banados2} (for previous theoretical study and support about this theory see~\cite{livro,born-infeld,Deser}).

The EBI theory is based on Palatini variational formulation, the metric and connection being treated as independent variables. It is indeed a special class of bimetric theories of gravity\footnote{See Refs. \cite{hassan,banados3} for discussions about the nature of bigravity theories such as stability, allowed parameter ranges and cosmological aspects.} and has already been used as an alternative to dark matter and dark energy, since the additional fields introduced can be interpreted as both components in different regimes~\cite{banados,banados3,setor,harko14,skordis} (see Refs. \cite{odintsov} for functional extensions of EBI theory). EBI gravity has also been investigated in issues such as black hole geometries~\cite{olmo}, structure and stability of compact stars~\cite{sham,harko2} and singularity-free cosmologies~\cite{banados2,avelino}, among many others.

The aim of our work is to investigate if the geometric mass, arising from the extra component of the field equations, can explain the well-known virial theorem mass discrepancy in clusters of galaxies. For this purpose we use the collisionless Boltzmann equation in the EBI gravity field equations and derive a generalized virial theorem. We find that the total mass of a cluster of galaxies in the framework of EBI gravity is the sum of its baryonic mass and a geometric mass $M_q$ which accounts for most of the mass of the cluster.

The work plan is as follows: In Section \ref{Born-Infeld}, based on the Palatini variational approach, we present an outline of the generalized field equations in EBI theory, while in Section \ref{Field-Equations} we derive the basic equations for a static spherically symmetric gravitational field. The generalized virial theorem is deduced in Section  \ref{Virial-Theorem}, and in Section  \ref{aplications} we present some astrophysical applications. The results and conclusions of our work are presented in Section \ref{Remarks}.


\section{Eddington-Born-Infield gravity}\label{Born-Infeld}


Let us consider the action coupled to an Eddington-Born-Infield theory as proposed by Ba\~{n}ados in \cite{banados}:
\begin{eqnarray}
 S=\frac{1}{16\pi G}\int d^{4}x\left[\sqrt{|g_{\mu\nu}|}\,R +\frac{2}{\gamma l^{2}}\sqrt{|g_{\mu \nu} -l^{2}K_{\mu \nu} |}\right]
 +\int d^{4}xL_{m}(\psi,g_{\mu \nu}),
 \label{action}
\end{eqnarray}
where $G$ is the Newton's gravitational constant, $R$ is the Ricci scalar, $\gamma$ is  a dimensionless parameter and $l$ is a scale.  $|A_{\mu\nu}|$ denotes the absolute value of the determinant for any tensor $A_{\mu\nu}$. The metric $g_{\mu\nu}$ couples to matter fields, and $K_{\mu\nu}$ is the Ricci tensor constructed solely from the connections $C^{\alpha}_{\mu \nu}$ and its derivatives:
\begin{equation} \label{C-connections}
K_{\mu\nu} = \partial_{\alpha}C^{\alpha}_{\mu\nu} - \partial_{\nu}C^{\alpha}_{\mu\alpha} + C^{\alpha}_{\alpha\beta}C^{\beta}_{\mu\nu}
- C^{\alpha}_{\beta\mu}C^{\beta}_{\alpha\nu}\,.
\end{equation}
The action (\ref{action}) is a functional of $g_{\mu\nu}$ and of the independent connections $C^{\alpha}_{\mu \nu}$, which should not be confused with the Levi-Civita connections $\Gamma^{\alpha}_{\mu\nu}$ of the metric $g_{\mu\nu}$. The matter Lagrangian density $L_{m}$ depends only on the metric $g_{\mu\nu}$ and the matter fields $\psi$.  Varying the action (\ref{action}) with respect to $g_{\mu\nu}$ and $C^{\alpha}_{\mu \nu}$ we obtain the field equations
\begin{eqnarray}
G_{\mu \nu} = 8\pi GT_{\mu \nu} - \frac{1}{l^2}\sqrt{\frac{q}{g}}g_{\mu\alpha}q^{\alpha\beta}g_{\beta\nu}
 \label{campo} \\
K_{\mu \nu} =\frac{1}{l^2}\left(g_{\mu\nu} +\gamma q_{\mu\nu}\right),
 \label{2campo}
\end{eqnarray}
where $q_{\mu\nu}$ is a new metric satisfying the metricity condition $D_{\rho}(\sqrt{q}q^{\mu\nu})=0$, $D_{\rho}$ is the covariant derivative built with the connections $C^{\alpha}_{\mu\nu}$ and $q$ is the determinant of $q_{\mu\nu}$ (see \cite{banados} for some details of calculation).
The connections are then given by
\begin{equation}
 C^{\alpha}_{\mu \nu}=\frac{1}{2}q^{\alpha \sigma}(\partial_{\nu}q_{\sigma \mu}+\partial_{\mu}q_{\sigma \nu}-\partial_{\sigma}q_{\mu \nu}).
 \label{conexao}
\end{equation}
Equation (\ref{campo}) is the modified Einstein field equations, the second term on the right hand side being the contribution from the EBI action. In this framework the dynamic of the spacetime is described by equations  (\ref{campo}) and (\ref{2campo}). It is worth mentioning that in EBI gravity, one of the metric, $g_{\mu\nu}$, couples to matter fields and has important physical meaning since the matter follow the geodesics given by connections built with $g_{\mu\nu}$. Thus it is called physical metric or matter metric. The second metric, denoted by $q_{\mu\nu}$, generates the symmetric connections $C^{\alpha}_{\mu\nu}$ which are used to build the tensor $K_{\mu\nu}$. The connection $C^{\alpha}_{\mu\nu}$ is the geometric connection because it determines the curvature of space-time, the matter fields are not coupled to the metric $q_{\mu\nu}$ (for studies on matter-gravity coupling and doubly coupling in a bimetric gravity see, e.g., Refs. \cite{yashar,lavinia}).


\section{Field equations for a system of identical and collisionless point particles}\label{Field-Equations}


Let us assume that the geometry of the cluster can be described by a time-oriented Lorentzian four-dimensional space-time manifold with spherical symmetry\footnote{For a discussion on deviations from this assumption and the effects of a triaxial structure see \cite{Defilippis}.}. The metrics $g_{\mu\nu}$ and $q_{\mu\nu}$ of an isolated spherically symmetric cluster are given respectively by
\begin{equation}
 g_{\mu\nu}dx^{\mu} dx^{\nu}=-e^{\nu(r)}dt^{2}+e^{\lambda(r)}dr^{2} + r^{2}d\theta^{2} + r^{2}\sin^{2}\theta d\phi^{2},
 \label{metricag}
\end{equation}
\begin{equation}
 q_{\mu\nu}dx^{\mu} dx^{\nu}=-e^{\eta(r)}dt^{2}+e^{\alpha(r)}dr^{2} + r^{2}d\theta^{2} + r^{2}\sin^{2}\theta d\phi^{2},
 \label{metricaq}
\end{equation}
where $\nu(r)$, $\lambda(r)$, $\eta(r)$ and $\alpha(r)$ are functions of the coordinate $r$ only.
The galaxies in the cluster are considered identical, collisionless point particles of mass $m$, and their space-time distribution is described by a distribution function $f_{B}(x^{\mu},u^{\mu})$, defined in the phase space, which obeys the general relativistic Boltzmann equation. The number of galaxies per unit volume $n(x^{\mu},u^{\mu})$ of phase space, as well as the energy-momentum tensor of the matter $T_{\mu\nu}$, are determined by the distribution function $f_{B}(x^{\mu},u^{\mu})$ as $n=\int f_Bdu$ and $T_{\mu\nu}= \int f_{B}mu_{\mu}u_{\nu}du$, respectively,
where $u_{\mu}$ is the 4-velocity of each galaxy and $du=du_{r}du_{\theta}du_{\phi}/u_{t}$ is the invariant volume element of the velocity space. Let $\langle u_{i}^{2}\rangle$ ($i=t, r, \theta, \phi$) be the average value of the square of the components of the 4-velocity. This average is defined by $\langle u_{i}^{2}\rangle \equiv (1/n)\int f_Bu_i^2du$. Then we have, for $i=t$ for instance, $\langle u_{t}^{2}\rangle = (1/n)\int f_Bu_t^2du = T_{tt}/mn$. The mass density of galaxies is $\rho = nm$, so we have $T_{tt}=\rho\langle u_{t}^{2}\rangle$. Note that although we considered all the galaxies having the same mass $m$, the density $\rho$ is not constant since it depends of the number density $n(x^{\mu},u^{\mu})$ in the phase space.  The components of the tensor $T_{\mu\nu}$ are represented in terms of an effective density $\rho_{eff}$ and an effective anisotropic pressure with components radial $p^{(r)}_{eff}$ and tangential $p^{(\perp)}_{eff}$, which are defined as~\cite{jackson}
\begin{equation}
 \rho_{eff}=\rho\langle u_{t}^{2}\rangle, \quad p^{(r)}_{eff}=\rho\langle u_{r}^{2}\rangle, \quad p^{(\perp)}_{eff}=\rho\langle u_{\theta}^{2}\rangle =\rho\langle u_{\phi}^{2}\rangle.
 \label{densidade}
\end{equation}
Taking into account this form of the energy-momentum tensor and the metrics given by (\ref{metricag}) and (\ref{metricaq}), the gravitational field equations (\ref{campo}) describing a cluster of galaxies in EBI gravity take the form
\begin{eqnarray}
 e^{-\lambda}\left(\frac{\lambda^{\prime }}{r}-\frac{1}{r^{2}} \right)+\frac{1}{r^{2}}= 8\pi G\rho\langle u_{t}^{2}\rangle + \frac{1}{l^{2}}\frac{e^{\eta/2}e^{\alpha/2}}{e^{\nu/2}e^{\lambda/2}}\frac{e^{\nu}}{e^{\eta}}
 \label{campo0}
\end{eqnarray}
\begin{eqnarray}
 e^{-\lambda}\left(\frac{\nu^{\prime }}{r}+\frac{1}{r^{2}} \right)-\frac{1}{r^{2}}= 8\pi G\rho\langle u_{r}^{2}\rangle -\frac{1}{l^{2}}\frac{e^{\eta/2}e^{\alpha/2}}{e^{\nu/2}e^{\lambda/2}}\frac{e^{\lambda}}{e^{\alpha}}
 \label{campo1}
\end{eqnarray}
\begin{eqnarray}
 \frac{e^{-\lambda}}{2}\left(\nu^{\prime \prime}+\frac{\nu^{\prime }}{r}-\frac{\lambda^{\prime }}{r} +\frac{\nu^{\prime 2}}{2}
 - \frac{\nu^{\prime }\lambda^{\prime }}{2}\right)= 8\pi G\rho\langle u_{\theta}^{2}\rangle  -\frac{1}{l^{2}}\frac{e^{\eta/2}e^{\alpha/2}}{e^{\nu/2}e^{\lambda/2}}
 \label{campo2}
\end{eqnarray}
\begin{eqnarray}
 \frac{e^{-\lambda}}{2}\left(\nu^{\prime \prime}+\frac{\nu^{\prime }}{r}-\frac{\lambda^{\prime }}{r} +\frac{\nu^{\prime 2}}{2}
 -\frac{\nu^{\prime }\lambda^{\prime }}{2}\right)= 8\pi G\rho\langle u_{\phi}^{2}\rangle
  -\frac{1}{l^{2}}\frac{e^{\eta/2}e^{\alpha/2}}{e^{\nu/2}e^{\lambda/2}},
 \label{campo3}
\end{eqnarray}
where $'= d/dr$ and $''= d^2/dr^2$.
By adding the gravitational field equations (\ref{campo0})-(\ref{campo3}) we obtain the following equation
\begin{eqnarray}
 e^{-\lambda}\left(  \frac{\nu^{\prime \prime}}{2} + \frac{\nu^{\prime}}{r}+ \frac{\nu^{\prime2}}{4}
 - \frac{\nu^{\prime}\lambda^{\prime}}{4}\right) = 4\pi G\rho \langle u^{2}\rangle
 +\frac{1}{2l^{2}} \frac{e^{\eta/2} e^{\alpha/2}}{e^{\nu/2}e^{\lambda/2}}\left(\frac{e^{\nu}}{e^{\eta}}-\frac{e^{\lambda}}{e^{\alpha}}-2\right),
 \label{camposoma}
\end{eqnarray}
where $\langle u^{2}\rangle=\langle u_{t}^{2}\rangle+\langle u_{r}^{2}\rangle+\langle u_{\theta}^{2}\rangle+\langle u_{\phi}^{2}\rangle$. The second group of equations (\ref{2campo}) for the EBI gravity gives us
\begin{eqnarray}
&&e^{-\alpha}\left(\frac{\eta^{\prime}}{r}-\frac{\eta^{\prime}\alpha^{\prime}}{4} +\frac{\eta^{\prime \prime}}{2} +\frac{\eta^{\prime2}}{4} \right)=-\frac{1}{l^{2}}\frac{e^{\nu}}{e^{\eta}}-\frac{\gamma}{l^{2}}\label{2camp0123}, \\
&&e^{-\lambda}\left(\frac{\alpha^{\prime}}{r}+\frac{\eta^{\prime}\alpha^{\prime}}{4} -\frac{\eta^{\prime \prime}}{2} -\frac{\eta^{\prime2}}{4} \right)=\frac{\gamma}{l^{2}}\frac{e^{\alpha}}{e^{\lambda}}+\frac{1}{l^{2}},\\
&&\frac{e^{-\alpha}}{2}\left(\frac{\alpha^{\prime}}{r}+\frac{2e^{\alpha}}{r^{2}} -\frac{\eta^{\prime}}{r} -\frac{2}{r^{2}} \right)=\frac{1}{l^{2}}+\frac{\gamma}{l^{2}},\\
&&\frac{e^{-\alpha}}{2}\left(\frac{\alpha^{\prime}}{r}+\frac{2e^{\alpha}}{r^{2}} -\frac{\eta^{\prime}}{r} -\frac{2}{r^{2}} \right)=\frac{1}{l^{2}}+\frac{\gamma}{l^{2}}.
\label{2campo0123}
\end{eqnarray}
Using these equations, the second term on the right hand side of equation (\ref{camposoma}) can be untangled supposing that $\eta(r)$ and $\alpha(r)$ are slowly varying functions of the coordinate $r$ ($\eta^{\prime}$ and $\alpha^{\prime}$ small), such that we can neglect quadratic terms. Thus we rewrite (\ref{camposoma}) as
\begin{eqnarray}
e^{-\lambda}\left(  \frac{\nu^{\prime \prime}}{2} + \frac{\nu^{\prime}}{r}+ \frac{\nu^{\prime2}}{4}
 - \frac{\nu^{\prime}\lambda^{\prime}}{4}\right) =4\pi G\rho \langle u^{2}\rangle
 - \frac{1}{2l\sqrt{\gamma}} \frac{\left(\frac{\eta^{\prime} + \alpha^{\prime}}{r}+ \frac{2}{l^{2}}\right)}{\left(\eta^{\prime\prime} + \frac{\eta^{\prime} - \alpha^{\prime}}{r}\right)^{1/2}}\,
 \label{campoarrumado}
\end{eqnarray}
so that in the right hand side only the second metric (\ref{metricaq}) appears.


\section{The virial theorem in Eddington-Born-Infield gravity}\label{Virial-Theorem}


Now we generalize the virial theorem to apply it to galaxy clusters which are described by the distribution function $f_B$. This function, however, obeys the relativistic Boltzman differential equation which must be integrated over the velocity space and then, in conjunction with the gravitational field equation (\ref{campoarrumado}), provides the relativistic virial theorem for EBI gravity.


\subsection{The relativistic Boltzmann equation}


The transport equation for the propagation of a collisionless system of particles in a curved Riemannian space-time is given by the relativistic Boltzmann equation \cite{jackson,lindquist}
\begin{equation}
 \left(p^{\alpha}\frac{\partial}{\partial x^{\alpha}}-p^{\alpha}p^{\beta}\Gamma _{\alpha \beta}^{i}\frac{\partial}{\partial p^{i}}\right)f_{B}=0,
 \label{boltzmann}
\end{equation}
where $p^{\alpha}$ is the 4-momentum of the particle, and $\Gamma^{i} _{\alpha \beta}$ ($i=1,2,3$) are the Christoffel symbols associated to the metric (\ref{metricag}) which, in the EBI gravity, is the metric that couples to matter fields~\cite{skordis}.
A simplification of the Boltzmann equation comes about if we introduce at any point $x$ of the space-time an appropriate orthonormal frame of tetrads $e^{a}_{\mu}(x)$ ($a= 0,1,2,3$)  satisfying the condition $g^{\mu \nu}e_{\mu}^{a}e_{\nu}^{b}=\eta ^{ab}$,  where $\eta ^{ab}$ is the Minkowski metric tensor. Any tangent vector $p^{\mu}$ at $x$ can be expressed as $p^{\mu}=p^{a}e_{a}^{\mu}$, which defines the tetrad components $p^{a}$. In the case of the spherically symmetric line element given by equation (\ref{metricag}) we introduce the following frame of orthonormal vectors~\cite{jackson}:
$e_{\mu}^{0}=e^{\nu/2}\delta_{\mu}^{0}$, $e_{\mu}^{1}=e^{\lambda/2}\delta_{\mu}^{1}$, $e_{\mu}^{2}=r\delta_{\mu}^{2}$, and $e_{\mu}^{3}=r \sin \theta \delta_{\mu}^{3}$.
 Now, let $u^{\mu}$ be the 4-velocity of a typical galaxy, satisfying the condition $u^{\mu}u_{\mu}= -1$, with tetrad components given by $u^{a}=u^{\mu}e^{a}_{\mu}$. The relativistic Boltzmann equation (\ref{boltzmann}) in tetrad components is given by
\begin{eqnarray}
u^{a}e_{a}^{\mu}\frac{\partial f_{B}}{\partial x^{\mu}}+\gamma _{bc}^{a}u^{b}u^{c}\frac{\partial f_{B}}{\partial u^{a}} =0,
\label{boltzmann3}
\end{eqnarray}
where the distribution function $f_{B}=f_{B}(x^{\mu},u^{a})$ and $\gamma^{a}_{bc}=e_{\mu;\nu}^{a}e_{b}^{\mu}e_{c}^{\nu}$ are the Ricci rotation coefficients~\cite{jackson}.
Here we assume that the only coordinate dependence of the distribution function is upon the radial coordinate $r$, so that equation (\ref{boltzmann3}) becomes
\begin{eqnarray}
 \left(u_r - \frac{u_{t}^{2}}{2}\frac{\partial \nu}{\partial r}- \frac{u_{\theta}^{2}+u_{\phi}^{2}}{r}\right)\frac{\partial f_{B}}{\partial r}  - \frac{1}{r}u_{r}\left(u_{\theta}\frac{\partial f_{B}}{\partial u_{\theta}}+u_{\phi}\frac{\partial f_{B}}{\partial u_{\phi}}\right)
\nonumber\\ - \frac{1}{r}e^{\lambda/2}u_{\phi}\left(u_{\theta}\frac{\partial f_{B}}{\partial u_{\phi}}
- u_{\phi}\frac{\partial f_{B}}{\partial u_{\theta}}\right)\cot \theta = 0.
\label{boltzmann4}
\end{eqnarray}
The spherical symmetry requires that the coefficient of $\cot\theta$ be zero, which implies that the distribution $f_B$ is a function of $r, u_{r}$ and $u^{2}_{\theta}+u^{2}_{\phi}$ only. Multiplying equation (\ref{boltzmann4}) by $mu_{r}du$, integrating over the velocity space and assuming that $f_{B}$ vanishes sufficiently rapidly as the velocities tend to $\pm \infty$, we obtain
\begin{equation}
r\frac{\partial}{\partial r}[\rho \langle u_{r}^{2}\rangle]+\frac{\rho}{2} [\langle u_{t}^{2}\rangle]r\frac{\partial \nu}{\partial r}-\rho[\langle u_{\theta}^2\rangle+\langle u_{\phi}^{2}\rangle-2\langle u_{r}^{2}\rangle]=0.
\label{bintegracaov}
\end{equation}
Now we multiply equation (\ref{bintegracaov}) by $4\pi r^{2}$, and integrate over the cluster radius to obtain
\begin{equation}
2K - \frac{1}{2}\int^{r}_{0}4\pi r^{3}\rho \langle u_{t}^{2}\rangle\frac{\partial \nu}{\partial r}dr=0,
 \label{bintegracaoa}
\end{equation}
where
\begin{equation}
K = \int^{r}_{0}2\pi \rho\,[\langle u_{r}^{2}\rangle + \langle u_{\theta}^{2}\rangle+\langle u_{\phi}^{2}\rangle]r^{2}dr
\end{equation}
is the kinetic energy of the cluster, which is a fundamental piece in the virial theorem.


\subsection{Geometric quantities}


In what follows we introduce some approximations that apply to test particles in stable motion around central fields and define our geometric quantities based on the $q_{\mu\nu}$ metric. Firstly, we assume that $\nu(r)$ and $\lambda(r)$ are small and slowly varying functions of the radial coordinate, so that in the left hand side of equation
(\ref{campoarrumado}) the quadratic derivative terms can be neglected. Secondly, we assume that the galaxies in the clusters have velocities much smaller than the velocity of the light, so that $\langle u_{r}^{2}\rangle\approx\langle u_{\theta}^{2}\rangle\approx\langle u_{\phi}^{2}\rangle\ll\langle u_{t}^{2}\rangle\approx1$. Thus, equations
(\ref{campoarrumado}) and  (\ref{bintegracaoa}) become, respectively
\begin{eqnarray}
 \frac{1}{2}\frac{1}{r^2}\frac{d}{dr}\left(r^{2}\nu'\right)=4\pi G\rho + 4\pi G\rho _{q},
 \label{camporeduzido}\\
 2K - \frac{1}{2}\int^{r}_{0}4\pi \rho\,r^{3}\,\nu'\,dr=0,
 \label{2cinetica}
\end{eqnarray}
where we have defined a geometric density $\rho_{q}$ as
\begin{equation}
4\pi G\rho_{q}=-\frac{1}{2l\sqrt{\gamma}} \frac{\left(\frac{\eta^{\prime} + \alpha^{\prime}}{r}+ \frac{2}{l^{2}}\right)}{\left(\eta^{\prime\prime} + \frac{\eta^{\prime} - \alpha^{\prime}}{r}\right)^{1/2}}.
 \label{roq}
\end{equation}
Comparing (\ref{camporeduzido}) and (\ref{roq}) we see that the second metric (\ref{metricaq}) contribution appears as an additional source term.
The total mass of the system inside a radius $r$ is given by
\begin{equation}
M(r)=\int^{r}_{0}4 \pi \rho\,r^{2}dr.
\end{equation}
Therefore, multiplying equation (\ref{camporeduzido}) by $r^{2}$ and integrating from $0$ to $r$ we obtain
\begin{eqnarray}
 G M(r)=\frac{1}{2}\,r^{2}\nu' - GM_{q}(r),
 \label{camporeduz}
\end{eqnarray}
where $M_q$ is the geometric mass of the cluster, inside a radius $r$, defined by
\begin{equation}
 M_{q}(r)= \int_{0}^{r}4\pi\rho_{q}(r)\,r^2dr\,.
 \label{massaq}
\end{equation}
We associate to this geometric mass a geometric potential energy defined as
\begin{equation}
\Omega_{q}= -\int_{0}^{R}\frac{GM_{q}(r)}{r}\,dM(r)\,,
 \label{omegaq}
\end{equation}
and a geometric radius $R_{q}$ of the cluster
\begin{equation}
 R_q= \frac{M_q^2}{\int^{R}_{0}\frac{M_{q}(r)}{r}\,dM(r)}.
 \label{rq}
\end{equation}
In terms of this quantities we have
\begin{equation} \label{omegarq}
 \Omega_q = - \frac{GM_q^2}{R_q}\,,
\end{equation}
a relation which will be useful in what follows.


\subsection{The virial theorem}


Finally we multiply equation (\ref{camporeduz}) by $dM(r)/r$ and integrating from $0$ to $R$, we obtain
\begin{equation}
 2K+\Omega+\Omega_{q}=0,
 \label{virial}
\end{equation}
which is the generalized virial theorem in EBI gravity. $\Omega_{q}$ is given by (\ref{omegaq}), $\Omega$ is the gravitational potential energy due to the baryonic mass, given by
\begin{equation}
\Omega = -\int_{0}^{R}\frac{GM(r)}{r}\,dM(r)\,,
 \label{omega}
\end{equation}
and we used equation (\ref{2cinetica}) for the kinetic energy. In order to translate the virial theorem (\ref{virial}) into an expression for
$M_{\mbox{\scriptsize{v}}}/M$, we introduce the virial radius $ R_{\mbox{\scriptsize{v}}}$, defined as~\cite{jackson}
\begin{equation}
 R_{\mbox{\scriptsize{v}}}= \frac{M^2}{\int^{R}_{0}\frac{M(r)}{r}\,dM(r)},
 \label{rv}
\end{equation}
such that $\Omega=-GM^2/R_{\mbox{\scriptsize{v}}}$. In addition, taking the virial mass $M_{\mbox{\scriptsize{v}}}$, defined as \cite{jackson}
\begin{eqnarray}
 2K=\frac{G M^2_{\mbox{\scriptsize{v}}}}{R_{\mbox{\scriptsize{v}}}}\,,
 \label{definicaomv}
\end{eqnarray}
and substituting these definitions in (\ref{virial}), taking into account (\ref{omegarq}), we rewrite the virial theorem as
\begin{equation}
 \frac{M_{\mbox{\scriptsize{v}}}}{M}= \sqrt{1+ \frac{R_{\mbox{\scriptsize{v}}}}{R_q} \left(\frac{M_q}{M}\right)^2}\,.
 \label{mv}
\end{equation}
For most of the observed galactic clusters, the relation $M_{\mbox{\scriptsize{v}}}/M>3$ is true. Therefore, the unity term can be neglected in (\ref{mv}) and the virial mass in EBI gravity can be approximated by
\begin{eqnarray}
 M_{\mbox{\scriptsize{v}}}\approx M_{q}\sqrt{\frac{R_{\mbox{\scriptsize{v}}}}{R_q}}.
 \label{mvaproximado}
\end{eqnarray}
According to equation (\ref{mvaproximado}), most of the mass in a cluster with mass $M_{tot}$ is in the form of the geometric mass $M_{q}$, so that
$ M_{q}\approx M_{tot}$. In other words, the ratio of the total mass and of baryonic mass is determined by a purely geometric quantity. The gravitational effects associated to the presence of the metric $q_{\mu \nu}$ could in principle be tested through gravitational lensing.


\section{Astrophysical applications} \label{aplications}



\subsection{Estimating the geometric mass for galactic clusters} \label{geometric_clusters_mass}


Clusters of galaxies are the largest and most massive self-gravitating bounded systems in the Universe. Although clusters are dynamically evolving, and deviations from hydrostatic and virial equilibrium must be expected, simulations from Gaussian random density fields predict surprisingly tight virial relation~\cite{mathiesen} (see also~\cite{evrard}). Here we intend to estimate the order of magnitude of the geometric mass, hence we choose the simplest relations in modeling clusters.

It is well known that inside clusters of galaxies there is a hot tenuous gas, observed by its emission in X-rays predominantly through thermal Bremsstrahlung~\cite{reiprich}. The following equation for the intracluster gas density $\rho_g$, known as $\beta$-model, provides a reasonably good description of the observational data~\cite{arnoud}:
\begin{equation}
 \rho _{g}(r)=\rho_{0}\left(1+\frac{r^{2}}{r_{c}^{2}}\right)^{-3\beta/2}
 \label{romodelo}
\end{equation}
where $r_{c}$ is the core radius, $\rho_{0}$ and $\beta$ are cluster-dependent constants. The mass $M_g(r)$ of the gas inside a radius $r$ is then given by
\begin{equation} \label{gas-mass}
M_g(r)= \int _{0}^{r}4\pi\rho_{g}r^{2}dr = 4\pi\rho_0\int _{0}^{r}\frac{r^2dr}{(1+r^2/r_c^2)^{3\beta/2}}\,.
\end{equation}
The total mass $M_{tot}(r)$ inside the cluster can be obtained as function of the gas density via Jeans's equation for a spherical system~\cite{binney}
\begin{equation}
\frac{1}{\rho_g}\frac{d}{dr}(\rho_g\langle u^{2}_{r}\rangle) +\frac{1}{r}\left(2\langle u^{2}_{r}\rangle -\langle u^{2}_{\theta}\rangle - \langle u^{2}_{\phi}\rangle\right)= -\frac{d\Phi}{dr}\,,
\label{mjeans}
\end{equation}
where $\Phi(r)$ is the gravitational potential.
Assuming that the gas in the cluster is isotropically distributed, we take
$\langle u^{2}_{r}\rangle = \langle u^{2}_{\theta}\rangle = \langle u^{2}_{\phi}\rangle$, and the pressure $P_g = \rho_g \langle u^{2}_{r}\rangle$.
Since the gravitational field inside the cluster is weak, we assume that the gravitational potential satisfies the Poisson equation
$\nabla^2\Phi(r)=4\pi G\rho_{\mbox{\scriptsize{tot}}}$, where $\rho_{\mbox{\scriptsize{tot}}}$ includes energy density of other forms of matter different from gas, like luminous matter, massive neutrinos, etc.
Besides, the observed X-ray emission from the intracluster gas is usually interpreted by assuming that the gas is in isothermal equilibrium~\cite{reiprich,arnoud}. Therefore, one may assume that the pressure $P_{g}$ of the gas satisfies the ideal gas equation of state $P_{g}=(k_{\mbox{\tiny{B}}}T_{g}/\mu m_{p})\rho_{g}$, where $k_{\mbox{\tiny{B}}}$ is Boltzmann's constant, $T_{g}$ is the gas temperature, $\mu \approx 0.61$ is the mean atomic weight of the particles in the gas, and $m_{p}$ is the proton mass. A first integration of the Poisson equation gives $d\Phi/dr=GM_{\mbox{\scriptsize{tot}}}(r)/r^2$. Using (\ref{romodelo}) and (\ref{mjeans}) we obtain the total mass profile inside radius $r$ as
\begin{equation}
 M_{\mbox{\scriptsize{tot}}}(r)=\frac{3k_{\mbox{\tiny{B}}}\beta T_{g}}{\mu m_{p}G}\,\frac{r}{1+ r_c^2/r^2}.
 \label{massacomro}
\end{equation}

On the other hand, the total mass of the cluster, according to the modified EBI gravity, consists of the sum of the baryonic mass (mainly the intra-cluster gas), and the geometric mass, so  that
\begin{eqnarray}
 M_{\mbox{\scriptsize{tot}}}(r)=4\pi\int _{0}^{r}(\rho _{g}+\rho _{q})r^{2}dr.
 \label{massato}
\end{eqnarray}
It follows that the total mass inside the radius $r$ satisfies the following mass continuity equation
\begin{eqnarray}
 \frac{dM_{\mbox{\scriptsize{tot}}}(r)}{dr}=4\pi r^{2}\rho _{g}(r)+4\pi r^{2}\rho _{q}(r).
 \label{massatotal}
\end{eqnarray}
Therefore we obtain the expression for the geometric density inside the cluster by using equations (\ref{romodelo}) and (\ref{massacomro})
\begin{eqnarray}
\rho _{q}(r) =\frac{3k_{\mbox{\tiny{B}}}\beta T_{g}}{4\pi G\mu m_{p}}\frac{(1+3r_{c}^{2}/r^{2})}{(1+ r_{c}^{2}/r^{2})^{2}}\,\frac{1}{r^2}
-\frac{\rho_{0}}{(1+r^{2}/r_{c}^{2})^{3\beta /2}}.
\label{roqtotal}
\end{eqnarray}
as well as the geometric mass:
\begin{equation} \label{geometric_mass}
M_q(r)= \frac{3k_{\mbox{\tiny{B}}}\beta T_{g}}{G\mu m_{p}}\,\frac{r}{1+ r_c^2/r^2} - 4\pi\rho_0\int_0^r\frac{r^2 dr}{(1+ r^2/r_c^2)^{3\beta/2}}\,.
\end{equation}
In order to estimate the geometric mass $M_q(r)$, we consider regions where $r\gg r_c$. In this approach the equations (\ref{roqtotal}) and (\ref{geometric_mass}) reduces to
\begin{equation} \label{roapp}
\rho _{q}(r) \approx\frac{3k_{\mbox{\tiny{B}}}\beta T_{g}}{4\pi G\mu m_{p}}\frac{1}{r^2}  - \rho_0\left(\frac{r_c}{r}\right)^{3\beta}
\end{equation}
and
\begin{equation}
M_{q}(r)\approx \left(\frac{3k_{\mbox{\tiny{B}}}\beta T_{g}}{\mu Gm_{p}}\right)r - \left(\frac{4\pi \rho_{0}r_{c}^{3\beta}}{3(1-\beta)}\right)r^{3(1-\beta)}
\label{mqcomroqintegrado}
\end{equation}
respectively. If we neglect the contribution of the gas in comparing with the gravitational effect of the geometric mass due to EBI gravity, then we approximate the above two equations by
\begin{equation}
\rho _{q}(r) \approx \left(\frac{3k_{\mbox{\tiny{B}}}\beta T_{g}}{4\pi G\mu m_{p}}\right)\frac{1}{r^2}
\label{rhoqaproximado}
\end{equation}
and
\begin{equation}
M_{q}(r)\approx \left(\frac{3k_{\mbox{\tiny{B}}}\beta T_{g}}{\mu Gm_{p}}\right)r\,.
\label{mqaproximado}
\end{equation}
In this case, an upper bound for the cutoff of $M_{q}(r)$ may be estimated if we consider the coordinate radius for which the decaying density profile (\ref{rhoqaproximado}) becomes equal to the mean energy density of the Universe $\rho_{\mbox{\scriptsize{univ}}}$. Let us name this radius $R_q^{cr}$. So, assuming $\rho_q(R_q^{cr})=\rho_{\mbox{\scriptsize{univ}}}=3H^{2}/8\pi G = 4.6975 \times 10^{-30}h_{50}^{2}$ g/cm$^{3}$~\cite{reiprich} where
$H=50h_{50}$ km/Mpc/s, and noting that $k_{\mbox{\tiny{B}}}T_{g}\approx$ 5 keV for most clusters, we obtain
\begin{eqnarray}
 R_q^{(cr)} & = &\sqrt{\frac{3\beta k_BT_g}{4\pi G\mu m_p\rho_{\mbox{\scriptsize{univ}}}}} \nonumber \\
&  = & 25.06\sqrt{\beta}\sqrt{\frac{k_{\mbox{\tiny{B}}} T_{g}}{\mbox{5 keV}}}\,h_{50}^{-1} \mbox{Mpc}.
\label{roqtotalaproximado1}
\end{eqnarray}
By using (\ref{roqtotalaproximado1}) we find that the total geometric mass corresponding to this radius is
\begin{equation}
 M_{q}^{(cr)}=13.7 \times 10^{15}\beta^{3/2}\left(\frac{k_{\mbox{\tiny{B}}} T_{g}}{\mbox{5 keV}}\right)^{3/2}h_{50}^{-1} \mbox{M}_{\odot}.
\label{mqaproximado1}
\end{equation}
Considering $\beta \approx 2/3$ and $k_{\mbox{\tiny{B}}}T_{g}=$ 5 keV~\cite{reiprich} we obtain $R_q^{(cr)}=20.46\,h_{50}^{-1} \mbox{Mpc}=15 \mbox{Mpc}$ and
$M_{q}^{(cr)}=7.5\times 10^{15}h_{50}^{-1} \mbox{M}_{\odot}=5.5\times 10^{15} \mbox{M}_{\odot}$ for $h_{50}^{-1}=50/67.8$, a value which is consistent with observations~\cite{reiprich}.


\subsection{Typical values for the virial mass and radius} \label{tipical}


Astrophysical observations, together with cosmological simulations, are generally interpreted in terms of a virialized part of the cluster.
The radii commonly used are either $R_{200}$ or $R_{500}$. These radii corresponds to a fixed density contrast $\delta\approx 200$ (or $\delta\approx 500$) as compared to the critical density of the universe $\rho_{\mbox{\scriptsize{univ}}}$. The corresponding masses inside this radii are defined as $M_{200}$ and $M_{500}$ and it is usually assumed that the virial mass of the cluster is $M_{V}=M_{200}$ (or $M_{V}=M_{500}$) and the virial radius $R_{V}=R_{200}$ (or $R_{V}=R_{500}$)~\cite{reiprich}.
In order to compare the predictions of EBI gravity with the observations we estimate the virialized mass $M_V(r=R_V)$  using (\ref{roapp})
\begin{equation}  \label{48}
\delta\times\rho_{\mbox{\scriptsize{univ}}}\times R^2_{200} = \frac{3\beta k_BT_g}{4\pi G\mu m_p} - \rho_0\left(\frac{r_c}{R_{200}}\right)^{3\beta}\,R^2_{200}\,,
\end{equation}
and taking into account that for $r\gg r_c$ (\ref{romodelo}) give us $\rho_0(r_c/R_{200})^{3\beta}=\rho_g(R_{200})$. We obtain
\begin{equation}  \label{R_200}
R_{200} = \sqrt{\frac{3\beta k_BT_g}{4\pi G\mu m_p\rho_{\mbox{\scriptsize{univ}}}(\delta + \delta_g)}} = \frac{R_q^{cr}}{\sqrt{\delta + \delta_g}}\,,
\end{equation}
where $\delta_g = \rho_g(R_{200})/\rho_{\mbox{\scriptsize{univ}}}$ gives the density contrast of the intracluster gas as compared to the critical density of the universe at the radius $R_{200}$. Using the value found above for $R_q^{cr}$ and taking $\delta_g=20$, we obtain
\begin{equation}  \label{R200}
R_{200} \cong 1.38\, h_{50}^{-1}\mbox{Mpc}, \quad \Rightarrow M_{200} \cong 5.0\times 10^{14} h_{50}^{-1} \mbox{M}_{\odot}.
\end{equation}
This value for $R_{200}$ is comparable to the analogous value $R_U^{200}\cong 3.51\, h_{50}^{-1}\mbox{Mpc}$ found in \cite{harko} for brane world models.
Taking $h_{50}^{-1}=50/67.8$, Eq. (\ref{R200}) give us $R_{200}\cong 1.0\,\mbox{Mpc}$ and $M_{200}\cong 3.6\times 10^{14} \mbox{M}_{\odot}$.
A similar calculation, using now $\delta_g = \rho_g(R_{500})/\rho_{\mbox{\scriptsize{univ}}}=50$,  give us
\begin{equation}  \label{R500}
R_{500} \cong 0.87\, h_{50}^{-1}\mbox{Mpc}, \quad \Rightarrow M_{500} \cong 3.2\times 10^{14} h_{50}^{-1} \mbox{M}_{\odot}.
\end{equation}
Observations show that the mass in the clusters of galaxies range from $10^{13}h^{-1}_{50}M_{\odot}$ to $10^{15}h^{-1}_{50}M_{\odot}$~\cite{reiprich,sharon}, therefore the value of the geometric mass obtained in the framework of EBI gravity is consistent with observations\footnote{The reader should be aware that in some of this references the virial masses are displayed in terms of $h_{70}^{-1} (=1.4h_{50}^{-1})$.}.


\subsection{Radial velocity dispersion in galactic clusters}


From the observational viewpoint the virial mass $M_{V}$ is determined from the study of the velocity dispersion of the stars and of the galaxies in the clusters. In terms of the velocity dispersion $\sigma_1$ the virial mass can also be expressed as \cite{Carlberg}
\begin{equation}
M_{V}=\frac{3\sigma_1^2}{G}\,R_{V}.
\label{mvirializada}
\end{equation}
As done earlier, let us assume that the velocity distribution in the cluster is isotropic, that is:
$\langle u^{2}\rangle= \langle u_{1}^{2}\rangle +\langle u_{2}^{2}\rangle +\langle u_{3}^{2}\rangle = 3\langle u_{1}^{2}\rangle = 3\sigma_r^2$, where $\sigma_r^2$ is the radial velocity dispersion and is related to $\sigma_1$ by $\sigma_r^2=3\sigma_1^2$.
Under this assumption, the radial velocity dispersion relation for clusters of galaxies in EBI gravity can be derived from equation (\ref{bintegracaov}), rewritten as
\begin{equation}
\frac{d}{dr}(\rho \sigma_{r}^{2})+\frac{1}{2}\rho \nu^{\prime}=0.
\label{jeans}
\end{equation}
A first integration of the equation (\ref{camporeduzido}) yields
\begin{equation}
\frac{1}{2}r^{2}\nu'=GM_{q}(r)+GM(r)+C,
\label{jeans1}
\end{equation}
where $C$ is an arbitrary constant of integration. Substituting (\ref{jeans1}) into (\ref{jeans}) we obtain the solution
\begin{equation}  \label{solucaojeans}
\sigma_{r}^{2}(r)=-\frac{1}{\rho}\int ^{r}_{0}\left[GM_{q}(r)+GM(r)+C\right]\frac{\rho(r)dr}{r^2}.
\end{equation}
As an example, let us consider a simple case in which the density $\rho$ of the matter inside the cluster has a power law distribution given by
\begin{equation}
\rho(r)=\rho_{0}r^{-n},
\label{roconhecida}
\end{equation}
where $\rho_{0}$ and $n$ are positive constants. The corresponding mass profile is $M(r)=4\pi \rho_{0} r^{3-n}/(3-n)$. Assuming for the geometric mass $GM_{q}(r)=q_{0}r$ where $q_{0}=3k_{B}\beta T_{g}/\mu m_{p}$, we obtain the following expressions for the velocity dispersion:
\begin{equation}
 \sigma_{r}^{2}(r)=q_{0}+\frac{C}{2r}-2\pi G\rho_{0}\,r\ln r +\frac{C_{1}}{\rho_{0}}\,r,
 \label{sigma1}
\end{equation}
for $n =1$;
\begin{equation}
 \sigma_{r}^{2}(r)=\frac{q_{0}}{3}-\pi G\rho_{0}\left(\ln r +\frac{1}{4}\right)\frac{1}{r^{4}}+ \frac{C}{4}\frac{1}{r} +\frac{C_{1}}{\rho_{0}}r^{3},
 \label{sigma3}
\end{equation}
for $n =3$; and
\begin{equation}
 \sigma_{r}^{2}(r)=\frac{q_{0}}{n}+\frac{2\pi G\rho_{0}}{(n-1)(3-n)}r^{2-n}+\frac{C}{n +1}\frac{1}{r}+\frac{C_{1}}{\rho_{0}}r^{n},
 \label{sigma}
\end{equation}
for $n\neq 1,3$, where we have introduced a new constant of integration $C_1$.
The equations (\ref{sigma1})--(\ref{sigma}) are the velocity dispersion profiles predicted by EBI gravity and it can be used to estimate the virial mass in clusters of galaxies. It should be noted that in equations (\ref{sigma1})--(\ref{sigma}) the velocity dispersion shows the same radial dependency as that obtained in the framework of metric $f(R)$ gravity by \cite{lobo} and for brane world models by \cite{harko}. Therefore, measurements of radial velocity dispersions alone can not distinguish these three gravity models. On the other hand, the corresponding equations obtained by \cite{sefi} in the Palatini formalism for $f(R)$ gravity, as well as those obtained by \cite{capozziello} and \cite{DGP} for hybrid metric-Palatini gravity and warped DGP-inspired $\textsl{L}(R)$ gravity respectively, have mismatch in at least one term. This can be used, together with observational data, to test the different predictions of these theories. In general, the observational data are fitted with these functions by using a nonlinear fitting procedure (see e.g.~\cite{Carlberg}).


\section{Discussions and final remarks}\label{Remarks}


Dark matter is a fundamental ingredient of the modern Cosmology, without which it seems impossible to explain the formation of structures in the Universe. However, although many efforts have been made so far, there is no observational evidence of non-gravitational interactions for this type of matter; accelerator and decaying experiments give no support for the physics upon which the dark matter hypothesis is based. This, indeed, rises doubts about its physical existence and, in turn, opens the possibility that the Einstein's theory of gravitation breaks down at some scale (see~\cite{baker} for development of quantitative procedures for comparing tests of modern gravity theories on all scales).

The recent proposal by Ba\~{n}ados {\it et al.}~\cite{banados,banados2,skordis}, based on the Eddington-inspired-Born-Infeld gravity~\cite{livro,born-infeld,Deser}, has attracted considerable interest lately as a theory capable of explaining some of the dark sector phenomena. In particular, this theory can challenge the necessity of including huge amounts of dark matter to explain the mass discrepancy in clusters of galaxies. In the present paper we analyzed this problem in the context of EBI gravity by deriving the untangled modified gravitational field equations (\ref{campoarrumado}). The extra terms, appearing in the modified gravitational field equations, induce an additional gravitational interaction, which in principle can account for the missing mass in clusters of galaxies. Using reasonable approximations for weak central fields, and
taking into account the collisionless Boltzmann equation, we have derived a generalized version of the virial theorem within the context of EBI gravity (see Eq. (\ref{virial})). The new virial mass is mainly determined by the geometric mass associated with the geometrical terms (see Eqs. (\ref{mv})  and (\ref{mvaproximado})), showing the existence of a relationship of proportionality between the virial mass and geometrical mass of the cluster. Using the simplest relations and assumptions in modeling clusters of galaxies, as well as the intra-cluster hot gas, we have estimated the order of magnitude of the geometric mass, showing that it is compatible with observations (see Eqs. (\ref{R200}) and (\ref{R500})). In order to compare our results with the approximations made in section (\ref{aplications}) we write Eq. (\ref{mvaproximado}) for $R_q =R_q^{(cr)}$, giving the relation $M_q^{(cr)}/M_V=\left(R_q^{(cr)}/R_V\right)^{1/2}$. In subsection (\ref{geometric_clusters_mass}) we estimated a critical radius $R_q^{(cr)}$ taking into account the mean energy density of the Universe $\rho_{\mbox{\scriptsize{univ}}}$ today, obtaining $R_q^{(cr)}=15$ Mpc. Thus, taking $R_V=R_{200}=1$ Mpc, we obtain $M_q^{(cr)}/M_{200}\approx 3.87$, a reasonable value since $M_q^{(cr)}$ is un upper bound for the cutoff of $M_q$. On the other hand, as a result of our approximations in subsection (\ref{tipical}) we obtained that $R_q^{(cr)}$ can also be given by $R_q^{(cr)}=(\delta +\delta_g)^{1/2}R_{200}=(200+20)^{1/2} R_{200}$ (see Eq. (\ref{R_200})). Using again Eq. (\ref{mvaproximado}) with this value of $R_q^{(cr)}$ we obtain now $M_q^{(cr)}/M_{200}\approx 3.85$, which is almost the same value calculated above. A similar calculation, now with $M_{500}$ (see Eqs. (\ref{R500})), give us $M_q^{(cr)}/M_{500}\approx 4.84$ in the two approximations (remember now $R_q^{(cr)}=(500 + 50)^{1/2}R_{500}$). The similarity of this values shows the consistency of our result (\ref{mvaproximado}) as well as the adequacy of the approximations made in section (\ref{aplications}), specially in subsection (\ref{tipical}).

Our study of the virial theorem, generalized in the context of EBI gravity, makes evident the existence of a geometric mass (gravitational interaction) which would not be excluded by astrophysical observations. This might be an efficient tool to test the viability of this class of modified gravity. Finally, recalling that the virial mass $M_{V}$ is obtained from the observational study of the velocity dispersions of the stars in the cluster, we have derived a relation for the radial velocity dispersion (\ref{solucaojeans}) in the context of EBI gravity, which can be used for estimating virial masses.


\acknowledgments
The authors thank J.S. Alcaniz for his useful comments and review of the manuscript as well as the anonymous referee for his/her valuable remarks.  N.S.S. acknowledges financial supports by Conselho Nacional de Desenvolvimento Cient\'{i}fico e Tecnol\'{o}gico (CNPq) and Coordena\c{c}\~{a}o de Aperfei\c{c}oamento de Pessoal de N\'{i}vel Superior (CAPES) - Brazil. J.S. acknowledges financial support by CNPq and technical support by Departamento de F\'{\i}sica Te\'{o}rica e Experimental (DFTE-UFRN).


 \end{document}